\documentclass[AMS,STIX2COL]{WileyNJD-v2}

\articletype{Research Article}%

\begin{document}

\title{An Initial Exploration of Bayesian Model Calibration for Estimating the Composition of Rocks and Soils on Mars}

\author[1]{Claire-Alice H\'ebert*}

\author[2]{Earl Lawrence}

\author[2]{Kary Myers}

\author[3]{James P. Colgan}

\author[4]{Elizabeth J. Judge}

\authormark{H\'ebert \textsc{et al}}

\address[1]{\orgdiv{Department of Applied Physics}, \orgname{Stanford University}, \orgaddress{348 Via Pueblo, Stanford, CA 94305, USA}}

\address[2]{\orgdiv{Statistical Sciences}, \orgname{Los Alamos National Laboratory}, \orgaddress{\state{New Mexico}, \country{USA}}}

\address[3]{\orgdiv{Physics and Chemistry of Materials}, \orgname{Los Alamos National Laboratory}, \orgaddress{\state{New Mexico}, \country{USA}}}

\address[4]{\orgdiv{Chemical Diagnostics and Engineering}, \orgname{Los Alamos National Laboratory}, \orgaddress{\state{New Mexico}, \country{USA}}}

\corres{*Claire-Alice H\'ebert. \email{chebert@stanford.edu}}

\abstract[Summary]{
The Mars Curiosity rover carries an instrument, ChemCam, designed to measure the composition of surface rocks and soil using laser-induced breakdown spectroscopy (LIBS). 
The measured spectra from this instrument must be analyzed to identify the component elements in the target sample, as well as their relative proportions. 
This process, which we call disaggregation, is complicated by so-called matrix effects, which describe nonlinear changes in the relative heights of emission lines as an unknown function of composition due to atomic interactions within the LIBS plasma. 
In this work we explore the use of the plasma physics code ATOMIC, developed at Los Alamos National Laboratory, for the disaggregation task. ATOMIC has recently been used to model LIBS spectra and can robustly reproduce matrix effects from first principles. The ability of ATOMIC to predict LIBS spectra presents an exciting opportunity to perform disaggregation in a manner not yet tried in the LIBS community, namely via Bayesian model calibration. However, using it directly to solve our inverse problem is computationally intractable due to the large parameter space and the computation time required to produce a single output. Therefore we also explore the use of emulators as a fast solution for this analysis. 
We discuss a proof of concept Gaussian process emulator for disaggregating two-element compounds of sodium and copper. The training and test datasets were simulated with ATOMIC using a Latin hypercube design. After testing the performance of the emulator, we successfully recover the composition of 25 test spectra with Bayesian model calibration. 
}


\maketitle

\section{Introduction}\label{intro}
One of the main scientific drivers of the Mars rover Curiosity is to determine whether Mars has ever been host to forms of life \cite{Curiosity}. 
ChemCam, one of the instruments on board, developed by Los Alamos National Laboratory and L'Institut de Recherche en Astrophysique et Plan\'etologie, is designed to record detailed data about the surface soil and rocks of the planet. 
The instrument uses laser-induced breakdown spectroscopy (LIBS) to measure the abundance of all chemical elements by firing a laser onto a small patch of rock or soil surface, producing a plasma. 
As the high-temperature plasma cools, it emits light that ChemCam records via a spectrometer and CCD camera, producing a detailed spectrum over a range of wavelengths. 
Intensity peaks in such a spectrum can be used to identify the presence and relative abundance of chemical species in the  sample of rock or soil. 
The absence or presence of certain elements could hold important evidence about the possibility of life on the rocky planet. 

Curiosity can obtain hundreds of these spectra every day, but analyzing them remains a slow manual process. 
To successfully analyze, or disaggregate\footnote{Note that we use the term \textit{disaggregation} for our work to allow for eventual consideration of other types of problems, such as estimating the devices drawing power from a household given a single measurement of the household's total power usage. The chemistry community uses the term \textit{calibration} for this task, which also introduces some vocabulary overloading with our use of Bayesian model calibration.}, each spectrum, we must answer two questions: what elements were present in the sample, and what are their relative proportions or abundances? 

An expert can often answer the first question, identifying component elements by the presence of signature peaks in the spectrum at specific wavelengths. 
Estimating the relative abundance of these constituent species, however, can be quite difficult due to atomic interactions within the plasma. 
Due to these interactions, the relative heights of emissions lines of two elements can have a nonlinear dependence on the relative abundance of those species. 
These \textit{matrix effects} --- where \textit{matrix} refers to the components of the target rather than to the mathematical concept --- complicate our disaggregation problem: the spectrum of a multi-species target is not simply the linear combination of the spectra for each individual element. 
These nonlinear effects pose significant challenges for disaggregation of LIBS data, particularly because we don't have a closed form expression of the effects. 

\citet{Clegg2009} introduced the use of multivariate approaches, such as partial least squares regression, to improve the ability to determine the elements present in a sample and estimate their relative proportions in the presence of matrix effects. More recently, scientists at Los Alamos National Laboratory adapted a first-principles plasma physics code called ATOMIC \cite{Magee2004} to provide a forward model for the emission from LIBS plasmas, including their matrix effects \cite{Colgan15, Colgan14, Judge2016}. The new existence of this forward model presents an exciting opportunity to explore the use of \textit{Bayesian model calibration} \cite{kennedy2001} to compute estimates, with associated uncertainties, of the components of a target and their relative abundances. That is, we want to solve the inverse problem to determine the input parameters for ATOMIC, which include the elements present and their proportions, that produce the simulated spectrum that is most like an unknown measured spectrum.

Bayesian model calibration in the context of high-dimensional outputs \cite{Higdon2008}, of which LIBS spectra are an example, has a successful history in a variety of scientific applications. For example, in materials science, these methods have been applied to estimate strength parameters of aluminum alloys in hydrodynamic shock experiments \cite{Walters2018}. 

We draw inspiration from \citet{Judge2016}, who demonstrated the impact of matrix effects on sodium peak heights in a simple two-element mix of sodium and copper measured with LIBS. The authors also demonstrated the ability of ATOMIC to replicate those effects from first principles. 
In particular, their experimental observations, supported by ATOMIC's theoretical calculations and modeling, showed that the sodium lines increased significantly in emission intensity as more copper was added to the target. 
This effect is explained by an increase in electron density, due to the copper, leading to increased recombination within the plasma. 

In this paper we will also use simplified two-element targets of sodium and copper to begin to explore our ability to use Bayesian model calibration with ATOMIC to perform disaggregation. As we look forward to more complex targets drawn from the large parameter space of all possible elements, and particularly in the context of the high data-collection rate of ChemCam on Curiosity, we recognize that using ATOMIC directly in this framework will be computationally intractable. We therefore also present our results building and evaluating \textit{emulators} to provide fast approximations to the computationally expensive ATOMIC runs. 

Emulators are well-established tools in the context of slow computer models and their use for modeling spectra has been demonstrated in the field of cosmology. 
Large N-body simulations of the universe are prohibitively expensive, and emulating the matter power spectrum of the universe on cosmological scales was shown to be an effective solution \cite{Lawrence2017}.
Although these spectra have very different physical origins from those of LIBS, these prior results provide some context for our efforts presented here. 

Our team's ongoing and preliminary work \cite{bhat2020} includes success at estimating the plasma temperature and density of simple, fixed compounds, taking into account a structured discrepancy between ATOMIC simulations and measured LIBS data.
While \citet{Judge2016} showed that ATOMIC can replicate the matrix effects in experimental data, the discrepancy study in \citet{bhat2020} and our own explorations showed some as yet unresolved challenges when comparing ATOMIC and measured LIBS spectra in the quantitative way required to support Bayesian model calibration. Therefore we focus here on the methodology of performing disaggregation in the presence of matrix effects in a simplified scenario, using a test set of simulated ATOMIC spectra rather than of measured spectra as would be our ultimate intent.
This allows us to demonstrate the feasibility of using modular Bayesian model calibration to perform disaggregation with LIBS data.

We present an overview of the ATOMIC simulations, inputs, and outputs in the following section. In Section \ref{methods} we outline the statistical framework of emulation and Bayesian calibration used in this work, and we discuss their application to simulation data in Section \ref{results}. We conclude with further discussion of the relevance and context of our results and future directions. 

\section{plasma simulations}\label{data}

Computer model calibration relies on a small number of runs of a high-fidelity  simulation, tiled across parameter space. 
This set of simulations can then be used to build emulators which can quickly approximate the computer model at new parameter settings. 
In the context of the analysis of LIBS data, the model of interest is the ATOMIC forward model, a general purpose plasma modeling and kinetics code. 
It was developed to simulate the emission spectra of chemical compounds using first principles theoretical atomic physics \cite{Magee2004}, in particular, emission or absorption spectra from plasmas either in local-thermodynamic equilibrium (LTE) or in non-LTE.

ATOMIC simulations require a few primary inputs: the temperature and density of the plasma, and a model describing the atomic structure and scattering data of the material(s) constituting the plasma. 
These last, which include quantities such as energy levels and transition probabilities, are generated from the Los Alamos suite of atomic physics codes \citep{Fontes15}. 
The results in the simulations discussed here were generated from the CATS code \cite{Cowan} with modifications made for plasmas generated from LIBS \cite{Colgan14}. 
For a given temperature and density, ATOMIC then models the emissivity of the plasma by computing its average ionization. 

\begin{figure*}[ht]
\centerline{\includegraphics[width=8in]{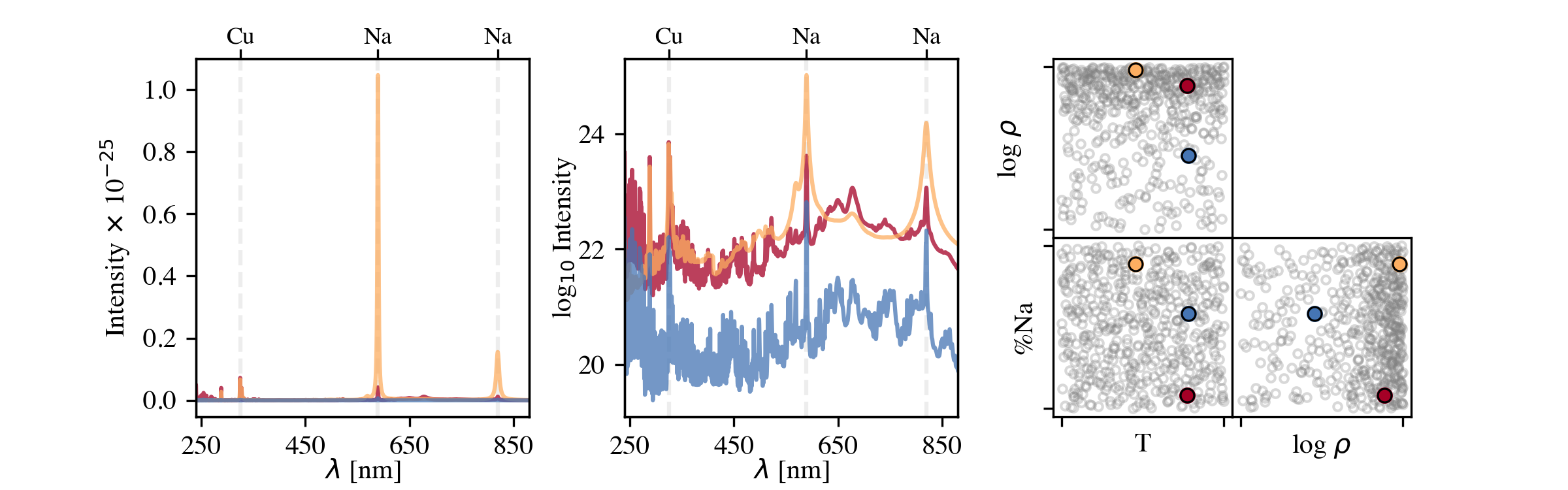}}
\caption{Three examples of ATOMIC simulations used for training along with the design for the training set. Left panel: three example training spectra, with intensity in units of power per volume per photon energy per unit solid angle. Middle: The same three examples plotted as log intensity, thereby reducing the orders of magnitude variation across the data.  Right panel: simulation input design for the 500 training spectra, scaled to the $[0,1]$ range for all parameters. The parameter settings for the three examples shown in the previous panels are highlighted in color. \label{dataexamples}}
\end{figure*}

As motivated above, the simulations discussed in this paper are of sodium copper (NaCu) plasmas with varying ratios of sodium (Na) to copper (Cu), inspired by \citet{Judge2016}. 
In addition, we vary the input plasma temperature $T$ and the mass density $\rho$ of the plasma. In particular, we run simulations over the following space of these three parameters:
\begin{enumerate}
\item plasma temperature $T$, in units electron volts (eV) and range $[0.5, 1.5]$
\item mass density $\rho$, in units of g/cm$^3$ and range $[-7,-4]$ on the log$_{10}$ scale  
\item composition of the plasma, defined as the proportion of one of the two elements in our compound. We have arbitrarily chosen to use the proportion of sodium in the plasma, $\%$Na, in the range $[0, 1]$. $\%$Cu can be retrieved by using $1 - \%$Na. 
\end{enumerate}

A few comments are in order about the third input parameter, $\%$Na. It is by solving the inverse problem for this parameter that we are doing disaggregation --- i.e., identifying the elements present and estimating their proportions. The validity of this particular parameterization, where $\%$Cu $= 1 - \%$Na, holds only because our simulations are run in an artificially simplified scenario without atmosphere, components of which, such as carbon and oxygen, would usually be present in the plasma in unknown quantities. Thus we know a priori that the only elements that could be present are sodium and copper. This enormously simplifies the first part of the disaggregation problem: identifying which elements are present. Our thinking is that success in this very constrained regime will establish a foundation for addressing the challenges faced by ChemCam on Mars, such as the presence of atmosphere and much larger sets of candidate elements. 

The output of the ATOMIC simulation is a spectrum with intensity as a function of wavelength for a particular set of the above inputs, i.e. a spectrum for a particular compound at some plasma density and temperature. The simulation provides intensity as power per volume per photon energy per unit solid angle. ATOMIC predictions do not account for any effects arising from the spectrometer, but do include matrix effects. The computation time for a single ATOMIC run depends on the chemical complexity of the compound of interest, and typically varies from minutes to hours on a high-performance computing system.

The ATOMIC simulations used in this analysis were selected using two Latin hypercube designs over the three input parameters: a training set of 500 simulations, and an independent test design of 25 points to which we added noise before analysis, as described in Section \ref{calibrations_results}. The training set parameter design, as well as a few of the resulting spectra, are shown in Figure \ref{dataexamples}. 
We have labeled the peak locations for two sodium peaks and one copper peak along the top wavelength axes, and will show these markers throughout this work when relevant. These are not the complete set of peaks for sodium or copper and are simply meant to provide some reference to the eye.

We note, in the left panel of Figure \ref{dataexamples}, the orders of magnitude over which the simulation output varies. This motivates a rescaling to the log scale, as shown in the middle panel, before statistical analysis, due to anticipated difficulty capturing small, yet potentially important, details in the presence of these large variations in amplitude. 
While this log transform is not standard practice in the spectroscopy community, exploratory analysis confirmed that emulators built for the spectra on their original scale performed significantly worse than the results presented here.
Additional motivations for modeling on the log scale are discussed in \citet{bhat2020}.

\section{Modular Bayesian calibration} \label{methods}

The aim of this work is to demonstrate the application of Bayesian calibration methods to the problem of disaggregating LIBS spectra --- identifying the elements present and estimating their proportions. We start with a brief overview of computer model calibration from \citet{kennedy2001}. In this context, computer model calibration entails estimating the input parameters of a computer model (here, ATOMIC) that most likely generated some given observed experimental or simulated data. 
A key assumption is that our observed data, $y$, are a noisy version of the simulator output at some unknown parameter setting $\theta$:
\begin{equation}\label{eq1}
    y = \eta (\theta) + \epsilon
\end{equation}
We will assume that the data $y$ have been centered and scaled according to the mean vector $\mu$ and scalar standard deviation $\sigma$ of the training data.

Following the well-established literature on calibration, we denote the ATOMIC computer model as $\eta(t)$, which takes a $p$-dimensional parameter vector $t$ as input to produce a LIBS spectrum.
Here, $p=3$ for the three input parameters described in Section \ref{data}.
The vector $\theta$ represents the parameter values that yield the model output $\eta$ that most closely resembles the observed data. Note that we are working here with $y$ and $\eta(t)$ as log scaled versions of the measured and modeled spectra. 

We estimate $\theta$ for a given observation $y$ by exploring the posterior $p(\theta | y)$ with Markov chain Monte Carlo (MCMC). 
This posterior is calculated, via Bayes rule, as the product of the data likelihood $p(y | \theta)$ and the parameter prior $p(\theta)$. 
We will first define the data likelihood, and the choice of prior is discussed below. 
Equation \ref{eq1} describes how the data are generated given the parameters $\theta$. The noise $\epsilon$ determines the sampling function for the data. Here we take  $\epsilon$ to be normally distributed with mean $0$ and variance $\Sigma_y$, which gives the following likelihood:
\begin{equation}\label{data_sampling}
    y | \theta \sim \mathcal{N} \left(\eta(\theta), \Sigma_y \right)
\end{equation}

This is in principle all we need (along with a prior) to perform MCMC. There is a significant computational challenge, though, as the ATOMIC model $\eta$ is slow to compute: given some vector $t$, evaluating $\eta(t)$ takes order of minutes or hours, rendering exploration via MCMC extremely slow. We will follow the standard approach to overcome this through use of an emulator: a statistical model that provides a fast approximation of the simulator output. 

In the following subsections we discuss first the approach of emulating the ATOMIC outputs using Gaussian processes, then details of Bayesian model calibration. 

\subsection{Gaussian process emulation} \label{emulation_methods}

As mentioned above, we define a statistical model to provide fast approximations of the slow ATOMIC outputs. This emulator will be some unknown function conditioned on a training set of $m$ simulator runs $\{\eta(t_1), ..., \eta(t_m)\}$ at fixed inputs $t_1, ..., t_m$. For simplicity, we scale these inputs such that $t\in [0,1]^p$. Before we describe the details of the emulator, we address a second computational bottleneck due to the high-dimensionality of the data. Each simulator output $\eta$ has $n_{\eta} = 32,000$ wavelength bins. Naive MCMC implementations might require the inversion of a $32,000m \times 32,000m $ matrix or $32,000$ $m \times m$ matrices at each step to calculate the posterior. The solution to this problem, as developed in \citet{Higdon2008}, relies on using dimensionality reduction to find a reduced set of basis vectors. This set of $n_{\eta}$-dimensional vectors $\{k_i, i=1,...,q\}$ describes the model for any input $t$ in the following way: 
\begin{equation}\label{decomposition}
    \eta(t) = \sum_{i=1}^q k_i w_i(t)
\end{equation}
where the weights $w_i(t)$ hold the dependence on the input $t$. We denote the number of components included in the reconstruction by $q$, with a maximum value of $m$, the size of training set. Typically smaller values $\sim10$ suffice for good performance. 
This formalism reduces the computational complexity of the problem enormously: we can now emulate and sample just $q\sim 10$ weights instead of $n_{\eta}\sim 10^5$ wavelength bins. 

We find this new basis via a singular value decomposition (SVD) of the training simulation matrix $X$. Each of the $m$ columns of $X$ holds a simulation of length $32,000$. The SVD factorization of $X$, in terms of orthogonal matrices $U, V$ and diagonal matrix $S$ of singular values of $X$, can be written:
\begin{equation}
    X = U S V^T = K W
\end{equation}
The second equality relates the SVD to Equation \ref{decomposition} in matrix form, with $K$ a column matrix of all the $k_i$s and the weights $w$ in row vector $W$. We have defined $K = US / \sqrt{m}$ and $W = V^T\sqrt{m}$. 

\begin{figure*}[ht] 
\centerline{\includegraphics[width=8in]{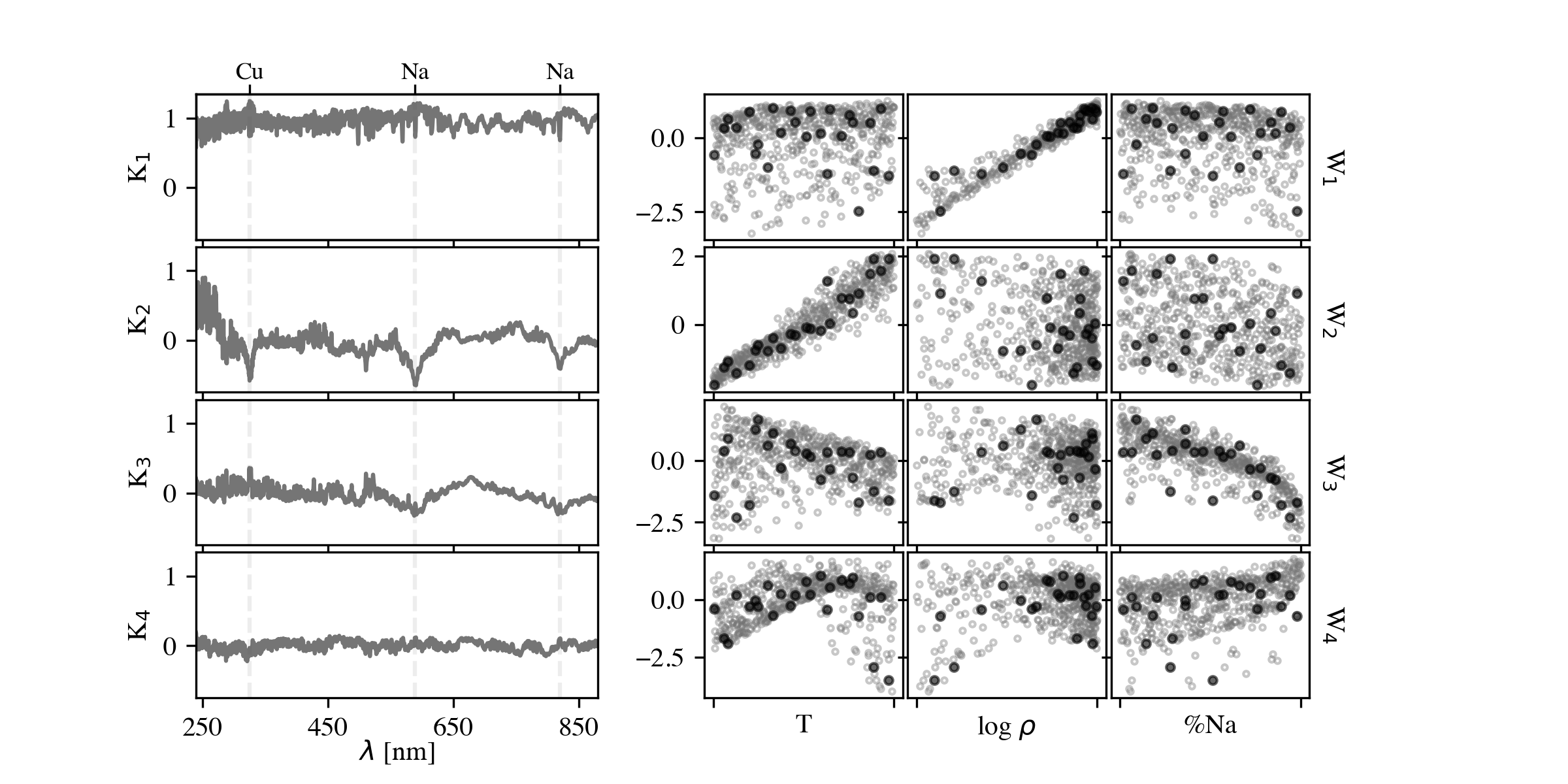}}
\caption{Exploring the singular value decomposition of the LIBS simulations. The left panel shows the first four basis components and the right panel shows the associated weights (in grey for each of the 500 training simulations; in black for  25 noise-added test examples) as functions of the three input simulation parameters, which are each scaled to lie in $[0,1]$. \label{svd_fig}}
\end{figure*}

In terms of the data described in Section \ref{data}, this decomposition results in a set of principal components $K$, each of which has length $n_{\eta} = 32,000$, that are common between all the spectra, and a set of weights which now hold the dependence on the input parameters and which vary between each spectrum. 
The first four of these principal components (PCs) are shown in Figure \ref{svd_fig} along with the associated weights for the 500 training and 25 test simulations.

Our ATOMIC emulator will take the form of Equation \ref{decomposition}, with $K$ calculated from the set of training simulations. The weights $w_i$ define a surface in parameter space for each component $i$ (see Figure \ref{svd_fig}): as is common in the literature, we will represent each of these by a Gaussian process (GP):
\begin{equation}
    w_i(t) \sim \mathcal{N}\left(0, \sigma_{wi}^2 R_i(t) \right)
\end{equation}
where $\sigma_{wi}^2$ is the marginal variance for weight $i$ and $R_i(t)$ is a correlation matrix with each entry given by the correlation function:
\begin{equation}\label{gpcov}
\text{corr}(t, t';l_i) = \prod_{j=1}^3 \exp \left(-\frac{|t_j-t_j'|^2}{2l_{ij}^2}\right)
\end{equation}
where $l_{ij}$ is the length scale hyperparameter for weight $i$ and parameter $j$.

Whereas it is common to perform the estimation of the GP hyperparameters concurrently with the calibration, we treat these in a modular way as in \cite{liu2009modularization}, fixing the hyperparameters by maximum likelihood estimation and keeping them fixed during the Bayesian calibration.
This approach greatly speeds up the estimation process because we do not need to rebuild and reinvert the covariance matrices at each step of the MCMC.
During calibration, we use the GPs to predict the value of each weight for parameter settings not present in the training simulations. 
The prediction of weight $i$ for some parameters $\theta$ can be found using properties of conditional normal distributions:
\begin{equation} \label{gp_predict}
    \hat{w_i}|\theta \sim \mathcal{N} \left(r_i(\theta)R_i^{-1}w_i, \sigma_{wi}[1 - r_i(\theta)^TR_i^{-1}r_i(\theta)] \right)
\end{equation}
Here $r_i(\theta)$ is the $m\times 1$ vector found by applying the GP covariance function in Equation \ref{gpcov} (with hyperparameters for weight $i$) to $\theta$ and the set of training simulation parameters $\{t_1,...,t_m\}$. 
The $m\times m$ matrix $R_i$ is the correlation matrix of the training parameters, and $w_i$ is the $m\times 1$ vector of training weights for principal component $i$. \\

\subsection{Model calibration}\label{calibration_methods}

We now describe the process of estimating the input parameters of the simulations via Bayesian model calibration.
As seen in Equation \ref{data_sampling}, we assume our data $y$ is given by a simulator run $\eta(\theta)$, with some fixed, diagonal covariance which can be parameterized by a precision $\lambda_y$: $\Sigma_y = \lambda_y^{-1} I$.
\begin{equation} \label{y_given_eta}
    y | \eta(\theta) \sim \mathcal{N} \left(\eta(\theta), \lambda_y^{-1}I \right)
\end{equation}

We have seen in the previous section how to express our data in a new basis of principal components $K$ and parameter-dependent weights $W$. 
The emulator output $\eta(\theta)$, given parameters $\theta$, is expressed in the new basis using the weights predicted by the GPs (see Equation \ref{gp_predict}): 
\begin{equation}\label{eta}
    \eta(\theta) = K\hat{w}(\theta)
\end{equation}
To cast a new observed spectrum $y$ into this basis, we use:
\begin{equation} \label{wobs}
    w_{obs} = \Tilde{K} y
\end{equation}
where $\Tilde{K} = (K^TK)^{-1}K^T$, and we define $w_{obs}$ as the vector of $q$ weights corresponding to the observation $y$. 

We use Equations \ref{y_given_eta}, \ref{wobs}, and \ref{eta} and properties of Gaussian distributions to find the sampling of the observed weights given $\hat{w}(\theta)$, the GP predictions:
\begin{align}
    w_{obs} | \hat{w}(\theta) &\sim \mathcal{N} \left(\Tilde{K} K\hat{w}(\theta), \Tilde{K} \lambda_y^{-1}I \Tilde{K}^T\right)\\
    &= \mathcal{N} \left(\hat{w}(\theta),  (\lambda_y K^TK)^{-1} \right)
\end{align}
Finally, since each predicted value $\hat{w}_i$ is drawn from a Gaussian process, these weights are normally distributed given parameters $\theta$, as in Equation \ref{gp_predict}. 
This means that the observed weights $w_{obs}$, given some parameter vector $\theta$, are given by:
\begin{equation}\label{likelihood}
    w_{obs} | \theta \sim \mathcal{N} \left(\mu_w, (\lambda_y K^TK)^{-1} + \Sigma_w \right),
\end{equation}
where $\mu_w$ is a vector with entry $i$ given by $r_i(\theta)_iR^{-1}w_i$, $i=1,...,q$ and $\Sigma_w$ is a $q\times q$ diagonal matrix with element $ii$ given by $\sigma_{wi} [1-r_i(\theta)^TR_i^{-1}r_i(\theta)]$, where the $i$ denotes that these are calculated from the covariance matrix with hyperparameters for weight $i$.
Equation \ref{likelihood} is the likelihood we will use for MCMC exploration.

\section{Results and discussion}\label{results}

We applied the methods above using the Python Scikit-learn implementation of Gaussian processes \cite{sklearn} and the Monte Carlo sampling in PyMC3 \cite{pymc3}. 
We chose a value of $q=15$ principal components in the emulator used to generate all the results discussed in this section. 
The SVD reconstruction with these 15 components explains $>99.995\%$ of the variance of the training set. In addition, the emulator and calibration errors did not significantly improve with added components beyond this point. 
Indeed, when a larger value is chosen for $q$, the additional weights added are less and less constraining for calibration since their associated sampling variances (see the likelihood in Equation \ref{likelihood}) increase with weight index. Adding more weights, then, is not expected to result in more accurate calibration results.

\begin{figure*}[ht!]
\centerline{\includegraphics[width=8in]{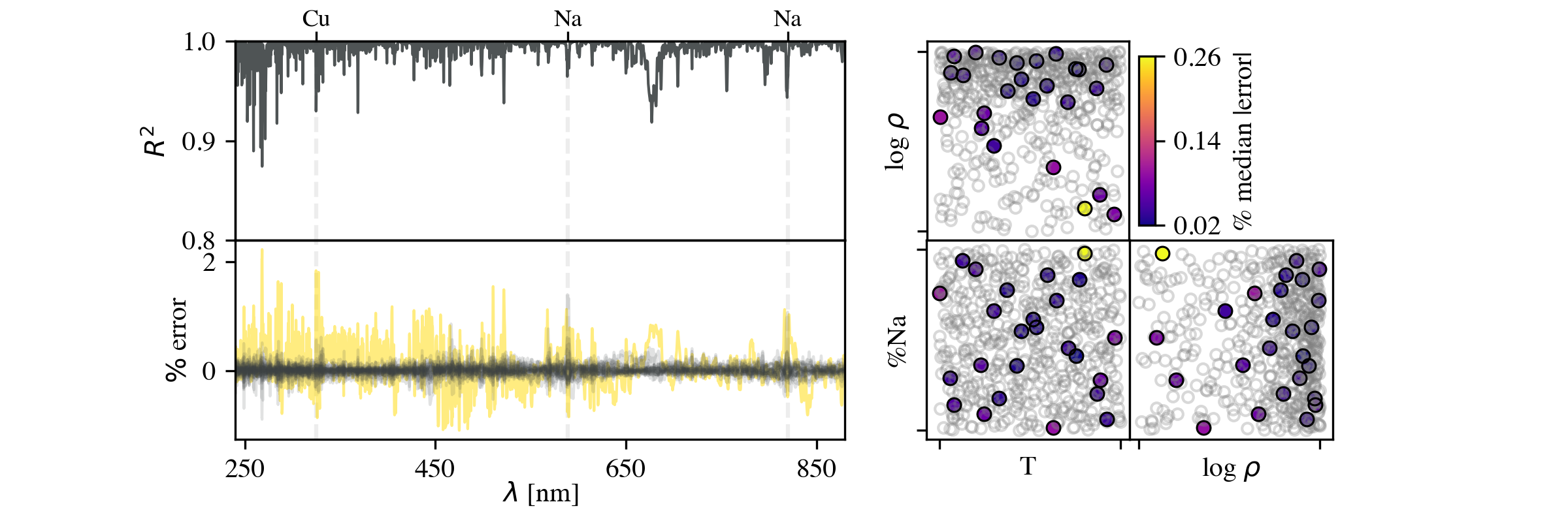}}
\caption{Summary of the emulator performance. Left: The $R^2$ and percent error as functions of wavelength (see Section \ref{emu_results} for details). The percent error for each of the 25 test runs is overlaid in the bottom panel. The yellow curve shows the percent error for the test run with the largest percent median absolute error, as shown in the right panel. Right: The test set design colored by percent median (over all wavelengths) absolute error of the emulator prediction for each point. The training design is shown in grey. \label{emulation_fig}}
\end{figure*}

Both emulation performance and calibration results are calculated based on a test set of 25 ATOMIC simulations run on a Latin hypercube design independent from the training set, to which we added noise. 

\subsection{Emulator performance} \label{emu_results}
Figure \ref{emulation_fig} summarizes the emulator performance for the settings listed at the start of this section. The performance is calculated on 25 test simulations.

We quantify emulator performance by two metrics: $R^2$ and percent error, both calculated point-wise for each wavelength modeled. We calculate these by:
\begin{itemize}
    \item $R^2 = 1 - \sigma^2_{res}/\sigma^2_{raw}$, where $\sigma^2_{raw}$ is the variance of the test simulations around their mean, and $\sigma^2_{res}$ is the variance of the residuals (emulator output - truth). 
    \item $\% \text{error} = 100 \times \frac{\sigma(\eta(\theta) - y)}{\sigma y + \mu}$, calculated for each test example and where $\mu, \sigma$ are the mean vector and scalar standard deviation of the training set.
\end{itemize}

Overall, these errors lie between $\pm 1\%$, with the exception of the test run plotted in yellow which we discuss below. We conclude that the emulator is performing well.

We expect that the information most important for our eventual goal of disaggregation should lie near the peaks associated with constituent elements. 
For example, the locations of some important sodium and copper peaks are indicated in Figure \ref{emu_results} by tick marks and vertical dashed lines. 
We see that the percentage error is greater at peak locations than at most other points along the wavelength axis.
The dataset contains a wide range of sodium concentrations, implying high data variance at these peak locations.
In absolute terms, the percent error shows that this variation is harder for the emulator to capture fully.
However, the spikes in $R^2$ value at those locations are not significantly worse than at many other wavelengths, indicating that the emulator is performing well relative to the variance of the data. 
In summary, the emulator is able to capture important variations in the data which we expect to be indicative of sodium concentration.

The right panel in Figure \ref{emu_results} shows the test design, each point colored according to the median percentage absolute error of the emulator predictions. 
The test run in yellow stands out as having the highest median error by far. Given its location on the very edge of parameter space for all three inputs, it is not surprising that the emulator has trouble with it.
The percentage error of this test run is highlighted in yellow in the left hand panel, and we will discuss it further below.

In summary, both the percentage errors and the $R^2$ values of the emulator predictions are indicative of strong performance, so we proceed with calibration.

\begin{figure*}[ht]
\centerline{\includegraphics[width=8in]{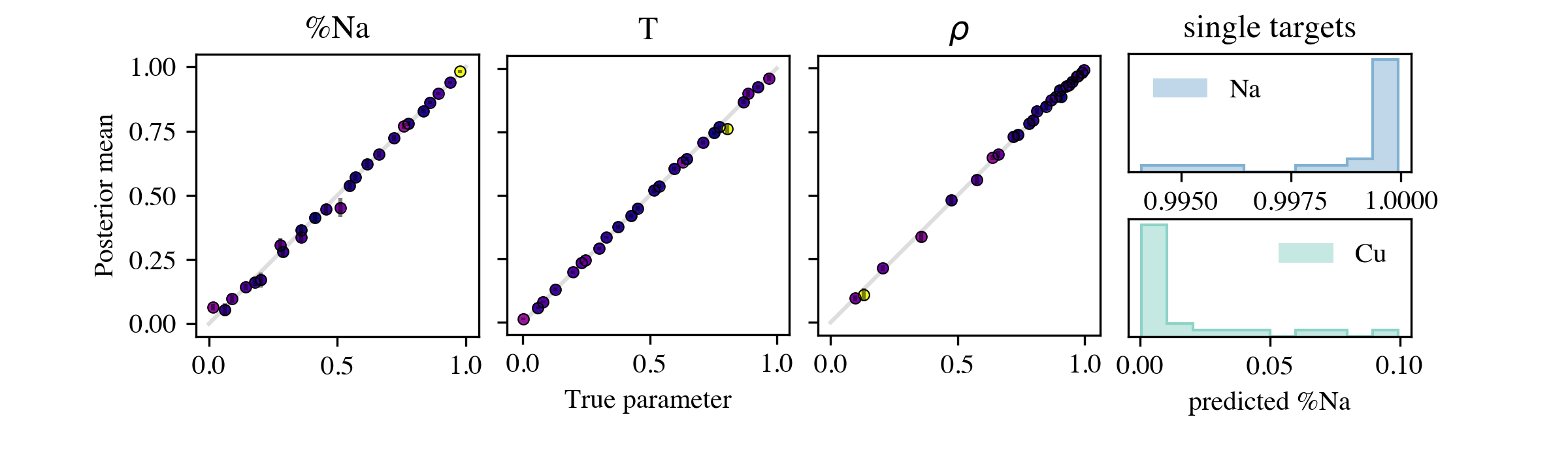}}
\caption{Summary of calibration results. The left three panels show the posterior mean of 15,000 MCMC samples for 25 noise-added test simulations, using the same color scheme as in the right panel of Figure \ref{emulation_fig} to indicate emulator errors. Panels show results for estimated \%Na, temperature $T$, and density $\rho$, compared to the true parameter value. Right: histogram of estimated \%Na from single target simulations (see Section \ref{calibrations_results}), for sodium-only targets in the upper panel and copper-only in the lower panel. \label{calibration_fig}}
\end{figure*}

\subsection{Calibration results}\label{calibrations_results}

The Bayesian calibration was performed for noise-added versions of the 25 test spectra, following Equation \ref{y_given_eta} with precision $\lambda_y=4$. 
In addition to the 25 two-element NaCu compounds with varying proportions of Na, we also ran the calibration for 50 spectra of single-element simulations, with the composition for 25 fixed at $100\%$ and 25 fixed at $0\%$ Na. 
These single-element simulations were run with the same input design for plasma temperature and density as the 25 two-element test simulations. 
During analysis these single-element spectra were treated identically to the two-element test examples.

The summary of these calibration results is shown in Figure \ref{calibration_fig}.
The left three panels of this figure show very encouraging calibration results for the two-target simulations, with the points colored according to the emulator errors as in the right panel of Figure \ref{emulation_fig}. The leftmost panel shows successful disaggregation for all 25 test spectra. 
Of particular interest is the test run with highest emulator error that we discussed earlier, indicated in yellow. Recall that its true parameters are the highest \%Na in the design, relatively high temperature $T$, and very low density $\rho$. 
We find that the calibration results for this challenging test run are on par with the rest of the test set.

Our single-element simulation results are summarized by the histograms in the last panel of \ref{calibration_fig}. Our success in recognizing pure Na and pure Cu demonstrates the ability of the emulator trained on the two-element NaCu simulations to correctly identify examples of single elements. We had had some concern that the difference between a target with a vanishingly small amount of one element and one with none at all would look like a discontinuity of some sort that would be difficult to capture with our emulators, and these results suggest that our emulators are robust to that difference. 

\begin{figure}[t]
\centerline{\includegraphics[width=3.25in]{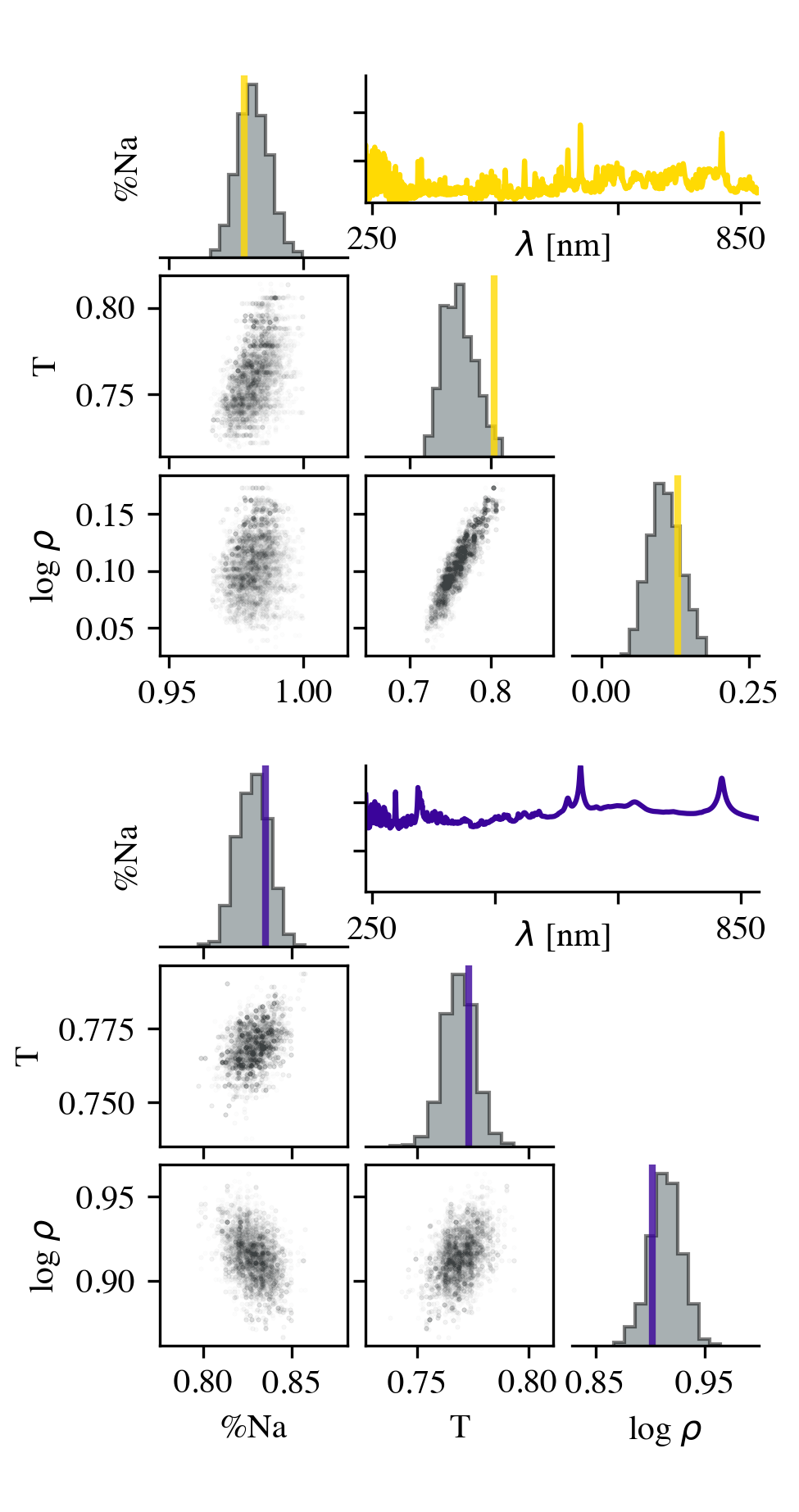}}
\caption{Example bivariate marginal distribution for two two-element test examples. 15,000 MCMC samples are shown for each. The log-scale test spectrum and true parameter values are shown, colored to match the associated emulator error from Figure \ref{emulation_fig}; for example, the upper triangle plot shows samples for the test example which demonstrated the highest emulator errors. The vertical range is the same for both spectra shown here, with ticks at log intensity of 20 and 22.5.
\label{mcmc_samples}}
\end{figure}

Figure \ref{mcmc_samples} shows bivariate marginal distributions of the MCMC samples alongside the log scaled test spectrum for two two-element test examples.
Results for the example with high emulator error discussed previously are displayed in the top panel, while a randomly selected example was chosen for the lower panel. Both spectra and true parameter values are shown in colors corresponding to their median absolute percent errors discussed in Section \ref{emu_results}. 
All three parameter combinations show some evidence of correlations between the samples, most notably between density and temperature in the upper panel.

\section{Conclusions}
In this work we show promising first results toward the goal of disaggregating LIBS data from ChemCam with Bayesian model calibration. The emulation and Bayesian calibration methods discussed here successfully perform disaggregation in the presence of matrix effects in noise-added simulations of two-element compounds. 

An important next step is to test this approach on measured LIBS data. 
There are a number of challenges to overcome before this can succeed. 
While ATOMIC simulations compare very well to experimental data on many metrics and for other applications \cite{Colgan15, Colgan14, Judge2016}, the wavelength-by-wavelength matching that drives the emulation in this work is not one of them. For example, we have found that the peak shapes, widths, and to some extent even locations, differ between ATOMIC outputs and measured LIBS spectra. 
These small effects have not been straightforward to accommodate with the methods presented here. 

In addition, LIBS spectra measured ``in the wild'' on Mars or even in a laboratory setting typically contain extraneous peaks that don't provide information about the target of interest. For instance, when measured in air, the plasmas include contributions from the atmosphere in unknown proportions that appear as peaks in the spectra. And laboratory targets of even simple two-element compounds like NaCu typically are formed using binders such as stearic acid that introduce still more peaks when measured via LIBS.  These extraneous components of the plasmas exacerbate matrix effects and complicate our analysis. A possible path forward that we are exploring is to limit the set of wavelengths in our analyses to those around expert-identified peaks of interest, rather than considering the entire spectrum shown in the work here.

In addition, instrument response and other effects should be accounted for when comparing simulations to measurements. Standard practice for ChemCam data is to correct for some of these effects, and we have begun exploring how these corrections impact our ability to perform disaggregation. 

While we continue those explorations toward comparing simulations with measurements as a separate line of research, we will build on our successes with simulated data that we presented here in order to provide a proof of principle for disaggregating increasingly complex multi-element targets.

\section*{Acknowledgments}
Research presented in this paper was supported by the Laboratory Directed Research and Development program of Los Alamos National Laboratory under project number 20180097ER. CAH was supported by the Department of Energy Computational Science Graduate Fellowship (DE-FG02-97ER25308).



\subsection*{Conflict of interest}

The authors declare no potential conflict of interests.



\nocite{}
\bibliography{main}%

\end{document}